\documentclass[journal=jacsat,manuscript=article]{achemso}

\usepackage{amsmath}
\usepackage{amssymb}
\usepackage{amsmath,amsthm,amssymb,amscd}
\usepackage{latexsym}
\usepackage{indentfirst}
\usepackage{subfigure}
\usepackage{graphicx}
\usepackage{bm}
\usepackage[version=3]{mhchem} % Formula subscripts using \ce{}

\author{Yunxin Zhang}
\email{xyz@fudan.edu.cn}
\affiliation{Shanghai Key Laboratory for Contemporary Applied Mathematics, Laboratory of Mathematics for Nonlinear Science, Centre for Computational Systems Biology, School of Mathematical Sciences, Fudan University, Shanghai 200433, China.}

\title[\texttt{achemso} demonstration]
{ATP binding to a multisubunit enzyme: statistical thermodynamics analysis}

\begin{document}
%%%%%%%%%%%%%%%%%%%%%%%%%%%%%%%%%%%%%%%%%%%%%%%%%%%%%%%%%%%%%%%%%%%%%
%% The manuscript does not need to include \maketitle, which is
%% executed automatically.  The document should begin with an
%% abstract, if appropriate.  If one is given and should not be, the
%% contents will be gobbled.
%%%%%%%%%%%%%%%%%%%%%%%%%%%%%%%%%%%%%%%%%%%%%%%%%%%%%%%%%%%%%%%%%%%%%
\begin{abstract}
Due to inter-subunit communication, multisubunit enzymes usually hydrolyze ATP in a concerted fashion. However, so far the principle of this process remains poorly understood. In this study, from the viewpoint of statistical thermodynamics, a simple model is presented. In this model, we assume that the binding of ATP will change the potential of the corresponding enzyme subunit, and the degree of this change depends on the state of its adjacent subunits. The probability of enzyme in a given state satisfies the Boltzmann's distribution. Although it looks much simple, this model can fit the recent experimental data of chaperonin TRiC/CCT well. From this model, the dominant state of TRiC/CCT can be obtained. This study provided a new way to understand biophysical processes by statistical thermodynamics analysis.
\end{abstract}

\section{Introduction}
In living cells, many enzymes consist of several subunits, each of which can bind and hydrolyze ATP by itself \cite{Adair1925, Braig1994, Noji1997, Riedel2007}. Because of the positive or negative cooperativity among the enzyme subunits, this ATP hydrolysis process is usually studied by Monod-Wyman-Changeux (MWC) model or  Koshland-Nemethy-Filmer (KNF) model \cite{Monod1965, Monod1966, Fersht1999, Kafri2003}. However, recent experimental measurements of the mammalian type II chaperonin TRiC/CCT, which consists of two ring-shaped cavities with lids composed of eight different subunits each, found that the bound ADP number presented on a single enzyme cannot be well described by the usual cooperative models, MWC model or KNF model \cite{Jiang2011}.

In experiments \cite{Jiang2011}, the stoichiometry of hydrolyzed ATP (in the form of either bound ADP or the transition state mimic, ADP$\cdot$AlFx) in a single TRiC enzyme is firstly measured, and then the hydrolyzed ATP number distribution is extracted.
In this study, we assumed that all the bound ATP molecules will be hydrolyzed, and so the number distribution of hydrolyzed ATP is just the number distribution of ATP bound to one enzyme.
As had been concluded \cite{Jiang2011}, many standard cooperative models can be ruled out by the measured shape of ATP number distribution.
In this study, we will not try to give a modified and so more complicated MWC or KNF model to fit the experimental data. On the contrary, we will present a simple thermodynamic model which is only based on Boltzmann's law.

\section{Results and discussion}

\begin{figure}
  \includegraphics[width=300pt]{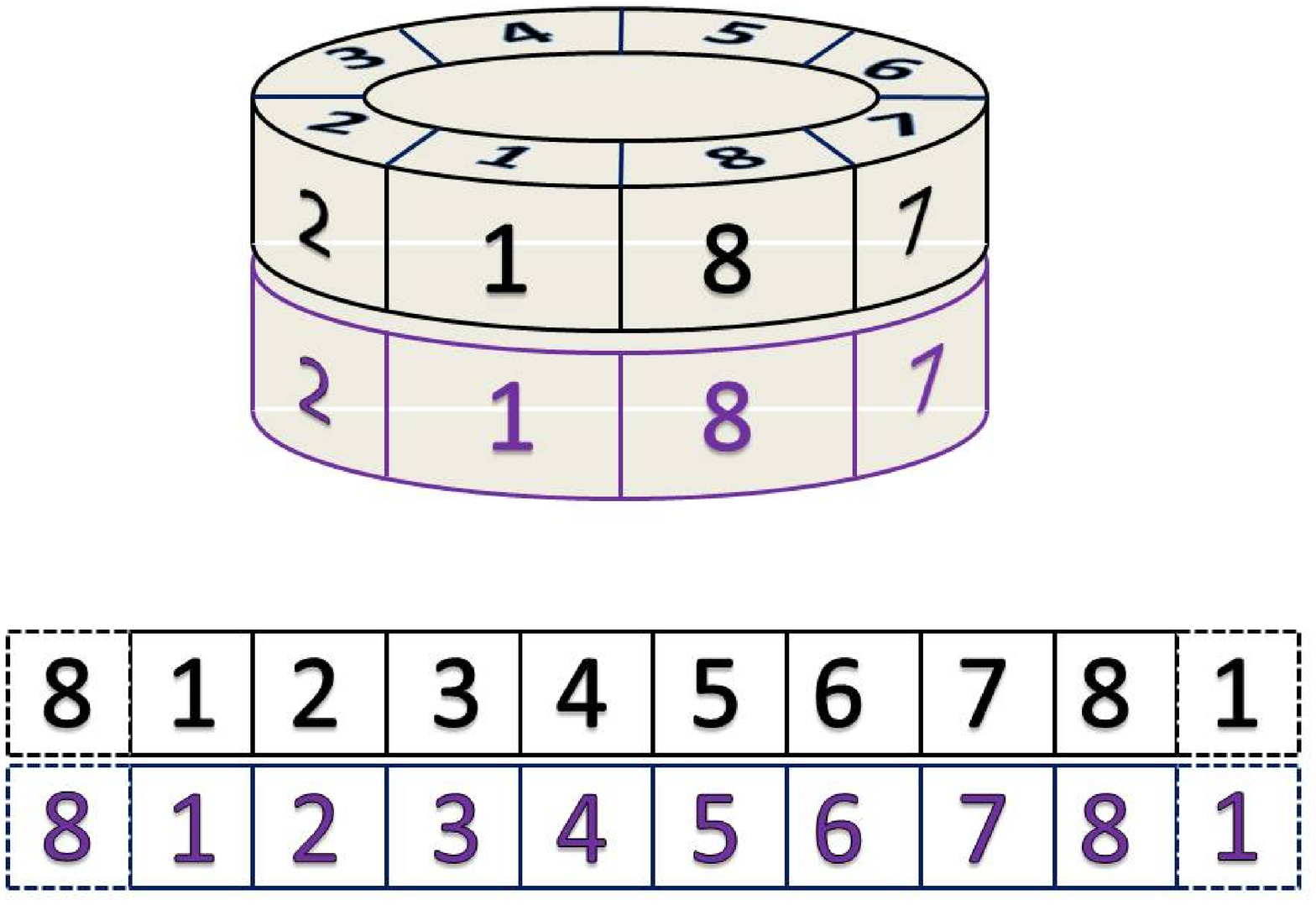}\\
  \caption{Schematic depiction of the chaperonin TRiC/CCT. TRiC consists of two stacked rings, each is made of eight CCT subunits, simply denoted by $1, 2, \cdots 8$ here.  Each CCT subunit might be in two states, 0 (empty) or 1 (ATP bound). So there are altogether $2^{16}=65536$ states. For convenience, the TRiC can also be regarded as two stacked chains with same period 8. The two CCT subunits with the same label number are said to be {\it mirror subunits} to each other.}\label{Figschematic}
\end{figure}

\begin{figure}
  \includegraphics[width=250pt]{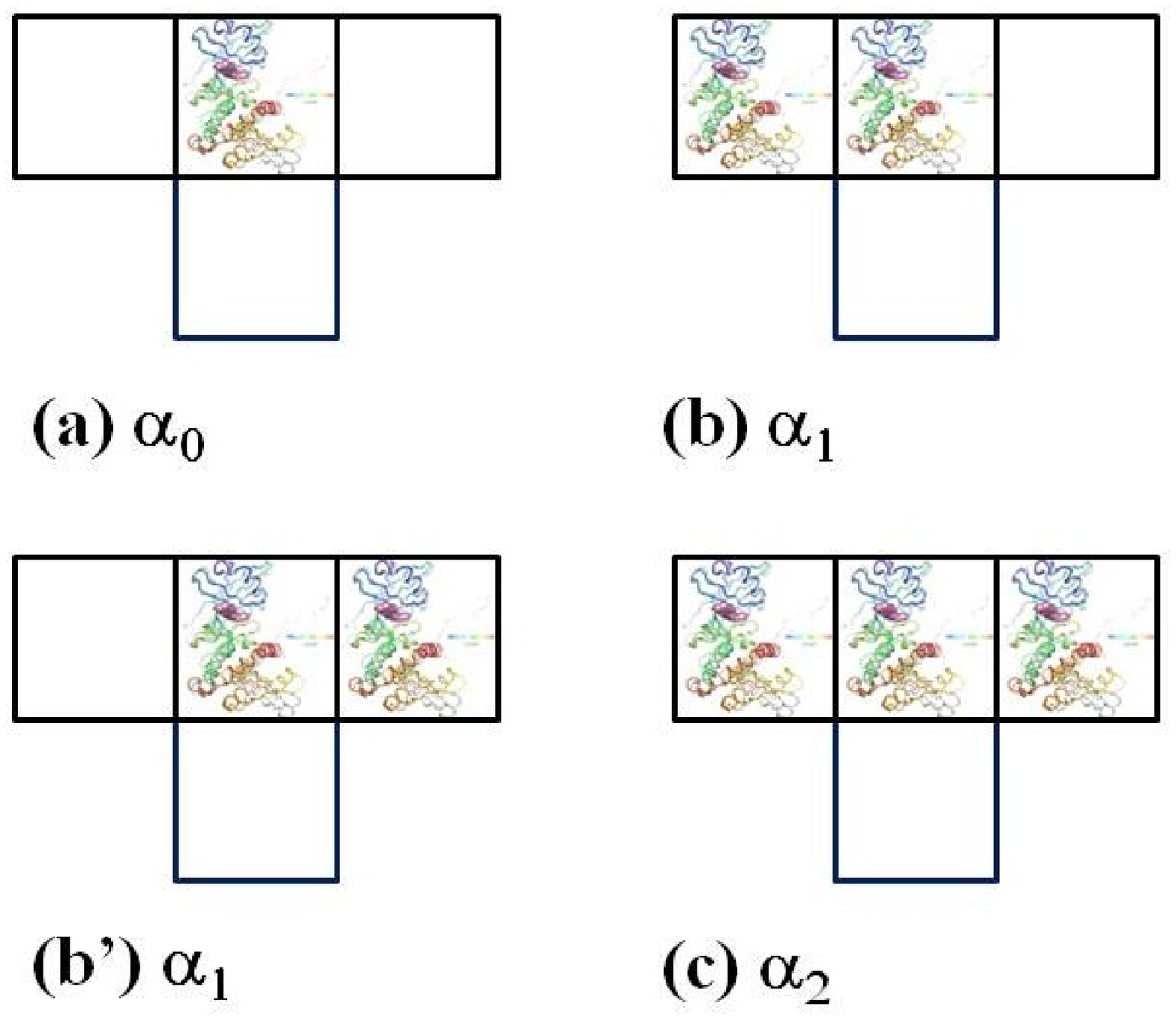}\\
  \includegraphics[width=250pt]{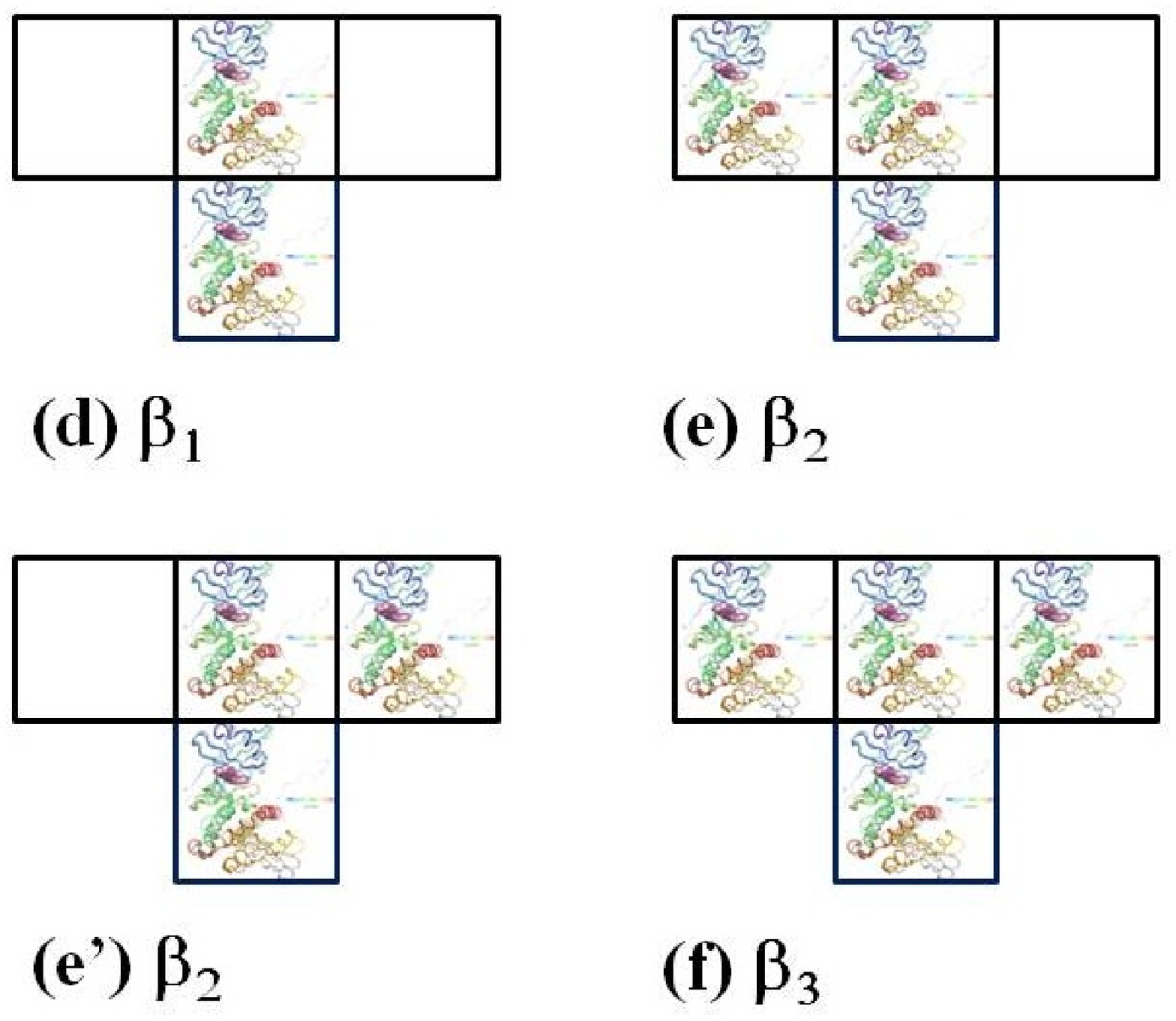}\\
  \caption{Potential of one ATP-subunit complex. We assume that, the potential changes of one ATP-subunit complex are the same when ATP binds to its left and right adjacent subunits [see (b) (b') and (e) (e')].
   %the effects of ATP binding to the left and right adjacent subunits to the potential of an ATP bound subunit are the same [see (b) (b') and (e) (e')],
   But they are different from the corresponding change when ATP binds to the {\it mirror subunit} [see (b) (d) and (c) (e): $\alpha_1\ne\beta_1$, $\alpha_2\ne\beta_2$]. The $\alpha_i, \beta_i$ below each figures are only potentials of the up-middle ATP-subunit complex, not the total potential of the four subunits. }\label{Figpotential}
\end{figure}

In our model, the enzyme TRiC/CCT is regarded as a double-ring with its two rings stack perfectly on one another (see \ref{Figschematic} \cite{Liou1997, Benito2004, Kabir2011}). Each ring consists of 8 CCT subunits which can bind and hydrolyze ATP by itself. The potential of each empty subunit (i.e. without ATP molecule) is denoted by $\epsilon$, and the potential of each free ATP molecule is denoted by $\gamma_1$. The (information) entropy of $K(\ge1)$ free ATP molecules is denoted by $\gamma_0\ln K$. For the ATP bound subunits, the potential of the ATP-subunit complex depends on the state of its adjacent subunits. For convenience, the two subunits which lie in different rings but be adjacent to each other, i.e. the two subunits with the same label number (see the schematic depiction in \ref{Figschematic}), are said to be {\it mirror subunit} to each other.
In the model, we assume that the two rings of TRiC are perfectly stacked, i.e. each subunit has and only has one adjacent subunit which lies in the different ring. For one ATP bound subunit, if its {\it mirror subunit} is empty and there are total $i$ ATP bound adjacent subunits, the potential of this ATP-subunit complex is denoted by $\alpha_i$, for $0\le i\le 2$. On the other hand, if its {\it mirror subunit} is ATP bound, and there are altogether $i$ ATP bound adjacent subunits, the potential of this ATP-subunit complex is denoted by $\beta_i$, for $1\le i\le3$ (see  \ref{Figpotential}).

Since each subunit of chaperonin might be empty (denoted by 0) or ATP bound (denoted by 1), for the double-ring chaperonin TRiC/CCT with $M=8+8$ subunits (generally, $M$ is between 14 and 18), there are altogether $2^M$ states, denoted by $s_k, k=1, 2, \cdots 2^M$ \footnote{For any $1\le k\le 2^M$, $s_k$ denotes the state $\{i_1i_2\cdots i_{M}\}$ of chaperonin which satisfies $\sum_{j=1}^{M}{i_j2^{j-1}}=k-1$, where $i_j=0$ or 1 for any $1\le j\le M$.}.
The total potential of the system with chaperonin in state $s_k$, including potential of the chaperonin subunits and potential of the $N$ ATP molecules which are bound to or are around enzyme chaperonin, is given as follows,
\begin{equation}\label{eq1}
\begin{aligned}
E_k=&\alpha_0N_0^{(1)}+\alpha_1N_1^{(1)}+\alpha_2N_2^{(1)}\cr
&+\beta_1N_0^{(2)}+\beta_2N_1^{(2)}+\beta_3N_2^{(2)}\cr
&+\gamma_0\ln(N-N_0-N_1-N_2)\cr
&+\gamma_1(N-N_0-N_1-N_2)\cr
&+\epsilon(M-N_0-N_1-N_2),
\end{aligned}
\end{equation}
where $N_i=N_i^{(1)}+N_i^{(2)}$ for $i=0, 1, 2$,
$N_i^{(1)}$ (or $N_i^{(2)}$) is the number of ATP bound subunits which have $i$ adjacent ATP bound subunits (in the same ring), and its {\it mirror subunit} is empty (or ATP bound).

Based on Boltzmann's law, the probability of chaperonin in state $s_k$ is
\begin{equation}\label{eq2}
\begin{aligned}
p_k=\frac{\exp(-E_k/k_BT)}{\sum_{i=1}^{2^M}\exp(-E_i/k_BT)}.
\end{aligned}
\end{equation}
Since the potential $E_k$ can be reformulated as follows
\begin{equation}\label{eq3}
\begin{aligned}
E_k=&(\alpha_0-\epsilon-\gamma_1)N_0^{(1)}+(\alpha_1-\epsilon-\gamma_1)N_1^{(1)}\cr
&+(\alpha_2-\epsilon-\gamma_1)N_2^{(1)}+(\beta_1-\epsilon-\gamma_1)N_0^{(2)}\cr
&+(\beta_2-\epsilon-\gamma_1)N_1^{(2)}+(\beta_3-\epsilon-\gamma_1)N_2^{(2)}\cr
&+\gamma_0\ln(N-N_0-N_1-N_2)+\gamma_1 N+\epsilon M,
\end{aligned}
\end{equation}
for a given system (i.e. with fixed $M$ and $N$), the expression \ref{eq2} of probability $p_k$ can be simplified as follows
\begin{equation}\label{eq4}
\begin{aligned}
p_k=\frac{\exp(-\hat E_k/k_BT)}{\sum_{i=1}^{2^M}\exp(-\hat E_i/k_BT)},
\end{aligned}
\end{equation}
where
\begin{equation}\label{eq5}
\begin{aligned}
\hat E_k=&\hat \alpha_0N_0^{(1)}+\hat \alpha_1N_1^{(1)}+\hat \alpha_2N_2^{(1)}\cr
&+\hat \beta_1N_0^{(2)}+\hat \beta_2N_1^{(2)}+\hat \beta_3N_2^{(2)}\cr
&+\gamma_0\ln(N-N_0-N_1-N_2),
\end{aligned}
\end{equation}
and
\begin{equation}\label{eq6}
\begin{aligned}
\hat \alpha_i=\alpha_i-\epsilon-\gamma_1, \quad
\hat \beta_i=\beta_i-\epsilon-\gamma_1.
\end{aligned}
\end{equation}
One can easily show that, the probability $P_n$ that one single chaperonin is bound by $n$ ATP molecules can be obtained as follows
\begin{equation}\label{eq10}
\begin{aligned}
P_n=\sum_{k\in I_n}p_k,\quad n=0 , 1, \cdots M,
\end{aligned}
\end{equation}
where
\begin{equation}\label{eq11}
\begin{aligned}
I_n=\left\{\sum_{i=1}^Ml_i2^{i-1}+1\left| \sum_{i=1}^Ml_i=n,\ l_i=0\ \textrm{or }1\right.\right\}.
\end{aligned}
\end{equation}
Obviously, there are $M\choose n$ numbers in set $I_n$, so intuitively, the maximum of probability $P_n$ will be attained at a number around $M/2$ [specially, if $\hat \alpha_i=\hat \beta_i=\gamma_0=0$, $P_n$ is binomial distributed, i.e. $P_n=2^{-M}{M\choose n}$. For further details, see the Langmuir adsorption model \cite{Langmuir1918, Volmer1925}].
The average number of ATP molecules bound to one Chaperonin is
\begin{equation}\label{eq12}
\begin{aligned}
\langle n\rangle=\sum_{n=1}^{M}nP_n=\sum_{k=1}^{2^M}n(k)p_k,
\end{aligned}
\end{equation}
where
\begin{equation}\label{eq13}
\begin{aligned}
n(k)=\sum_{i=1}^Ml_i,\quad\textrm{for\ } k=\sum_{i=1}^Ml_i2^{i-1}+1.
\end{aligned}
\end{equation}

\begin{figure}
  \includegraphics[width=300pt]{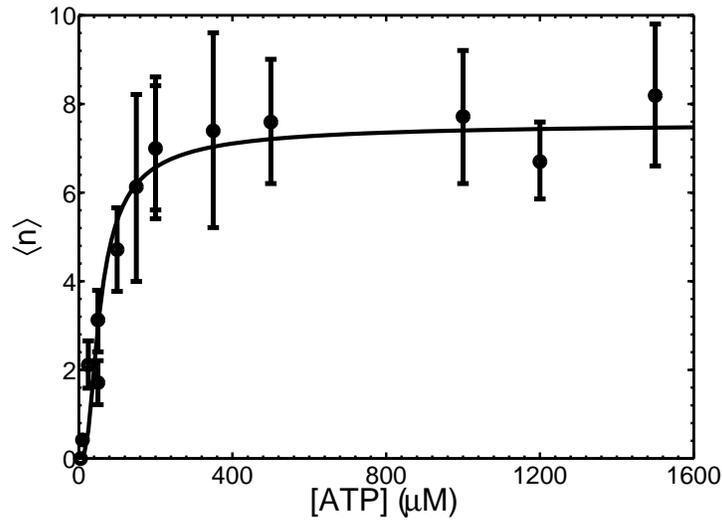}\\
  \caption{The statistical thermodynamics model prediction [solid curve, see  \ref{eq12}], and experimental data (solid dots) measured by Jiang {\it et al} in \cite{Jiang2011}, of the average ATP number $\langle n\rangle$ bound to one single chaperonin TRiC/CCT. In dilute ATP solutions, $\langle n\rangle$ increases rapidly with [ATP], and in saturating ATP solutions, $\langle n\rangle$ tends to 8.  See \ref{table1} for the parameter values used in the theoretical calculations. }\label{FigaverageN}
\end{figure}

\begin{figure}
  \includegraphics[width=310pt]{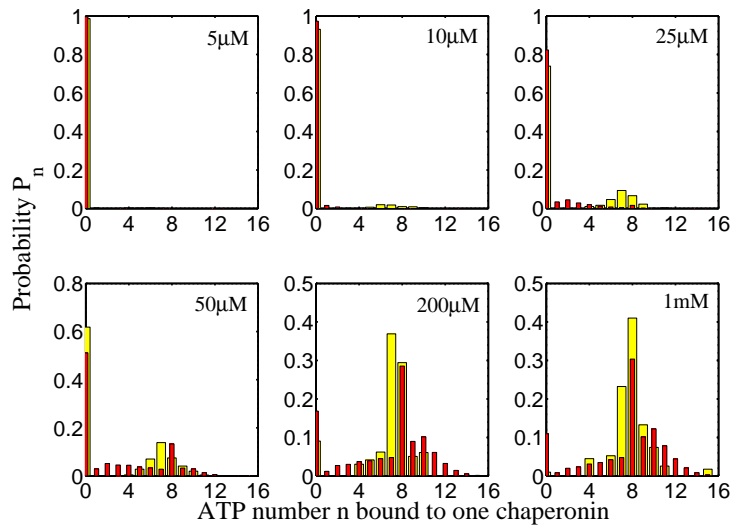}\\
  \caption{The model prediction [narrow bar, see \ref{eq10}] and experimental data (wide bar, measured by Jiang {\it et al} in \cite{Jiang2011}) of the probability $P_n$ of bound ATP number in one single chaperonin TRiC/CCT. For high [ATP], $P_n$ gets its maximum at $n=8$.  See \ref{table1} for the parameter values used in the theoretical calculations. }\label{FigprobabilityN}
\end{figure}

\begin{figure}
  \includegraphics[width=150pt]{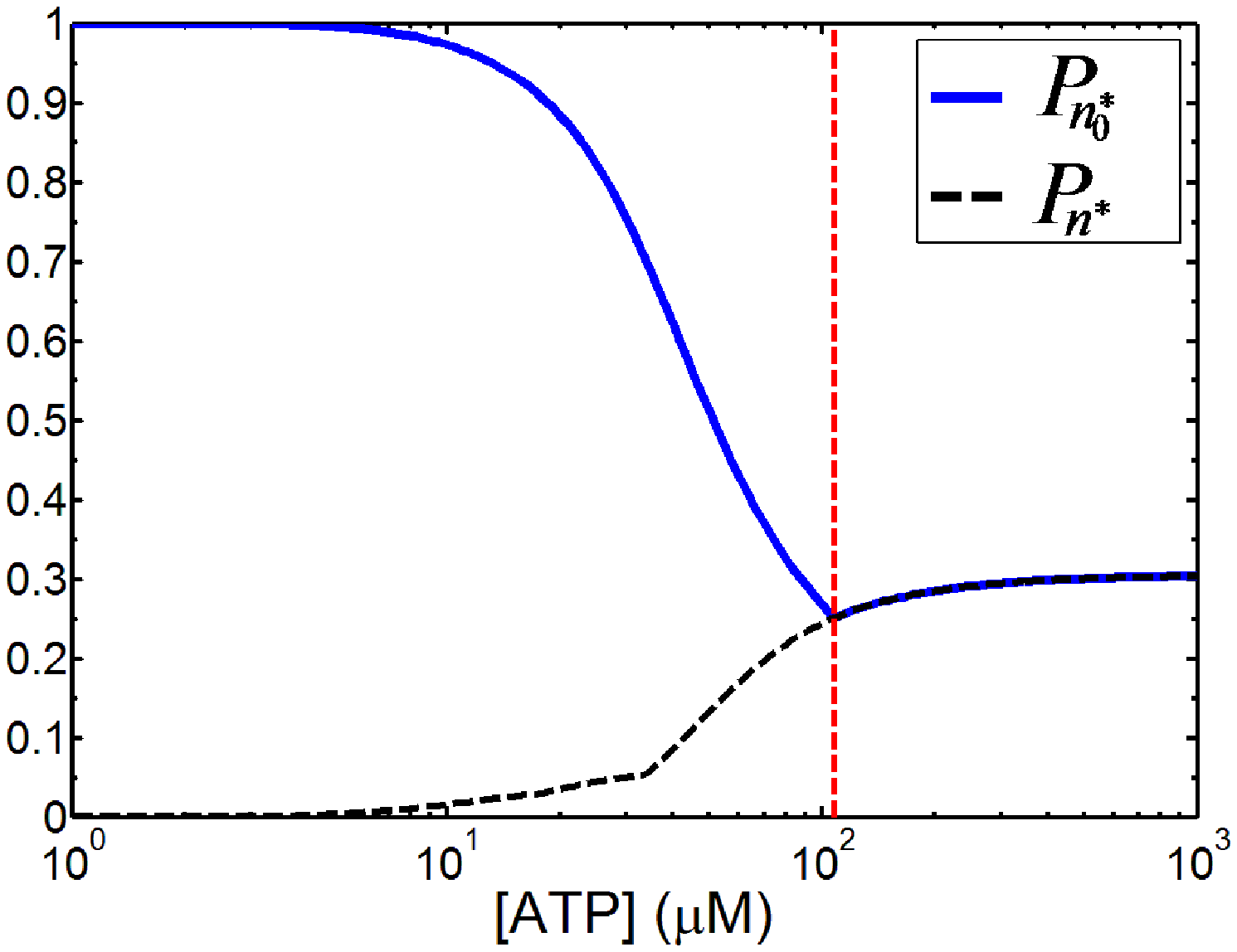}\includegraphics[width=150pt]{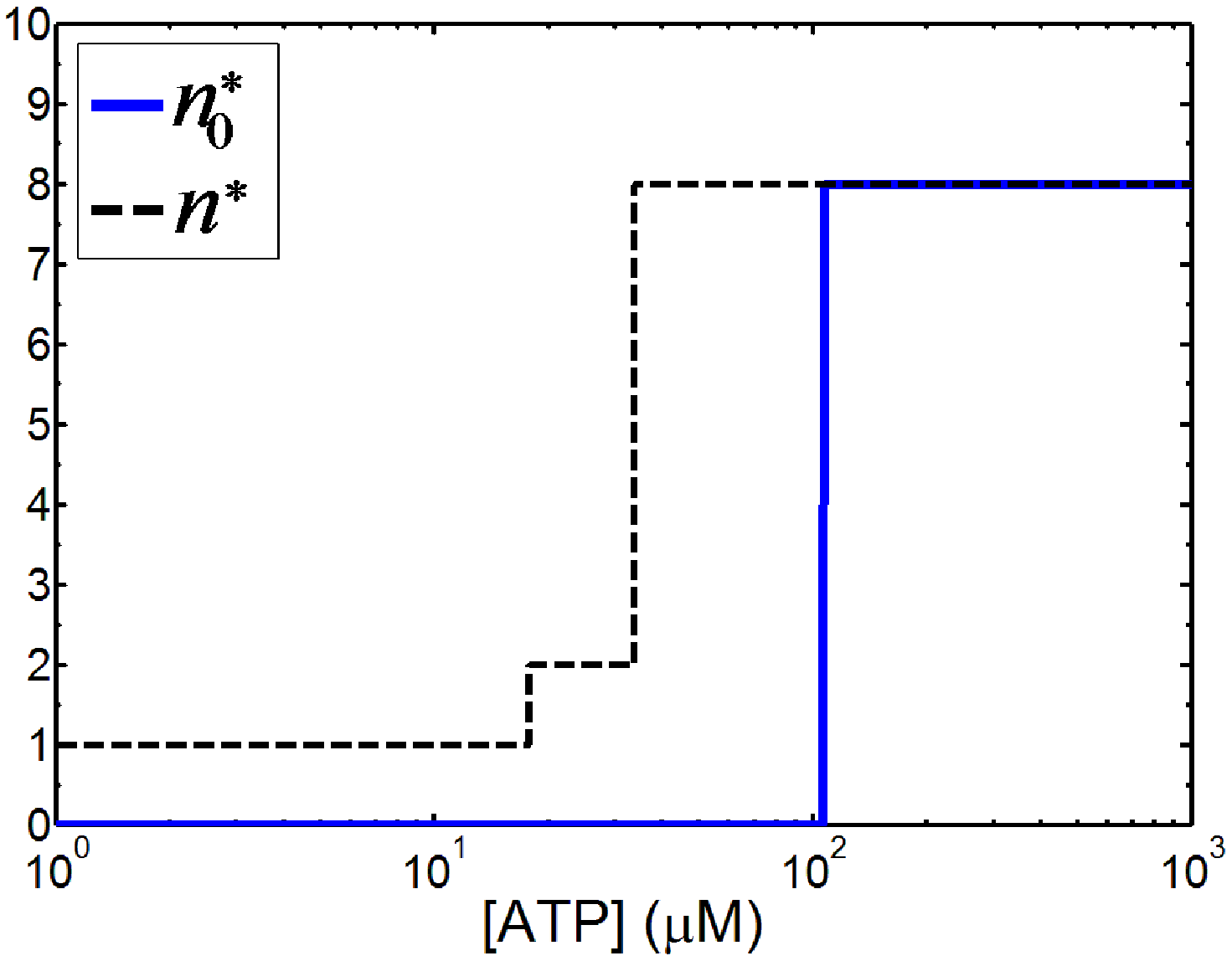}\\
  \caption{The maximal values $P_{n^*_0}$ and $P_{n^*}$ of probability $P_n$ of chaperonin TRiC/CCT, and the corresponding ATP binding number $n^*_0$ and $n^*$. Where $P_{n^*_0}=\max_{0\le n\le 16}P_n$ and $P_{n^*}=\max_{1\le n\le 16}P_n$. The critical [ATP] value in the left figure is about 107 $\mu$M. See \ref{table1} for the parameter values used in the theoretical calculations. }\label{FigmaxProb}
\end{figure}

With the model parameters listed in \ref{table1}, theoretical predications of average value $\langle n\rangle$ and probability $P_n$ are plotted in \ref{FigaverageN} and \ref{FigprobabilityN}, which imply that this simple statistical thermodynamics model can fit the experimental data of chaperonin TRiC/CCT well \cite{Jiang2011}. If chaperonin TRiC/CCT is incubated in low [ATP] solutions, the average value $\langle n\rangle$ increases rapidly with [ATP], but in saturating ATP solutions it tends to a constant 8.
For convenience, we denote the state of chaperonin TRiC/CCT by $[i_1i_2\cdots i_7i_8][i'_1i'_2\cdots i'_7i'_8]$, and define $P_{n^*}=\max_{0\le n\le 16}P_n$, $P_{n^*_0}=\max_{1\le n\le 16}P_n$. The plots in \ref{FigmaxProb} indicate that, for low ATP concentration, the dominant state is $[0\cdots0][0\cdots0]$, but for intermediate and high ATP concentration ([ATP]$>118\mu$M) the chaperonin TRiC/CCT most likely binds $n=8$ ATP molecules. By further numerical calculations, we found that for high [ATP], the dominant state of TRiC/CCT is $[11111111][00000000]$ or $[00000000][11111111]$. One also can see from the figures that, although in saturating ATP solutions the average value $\langle n\rangle$ tends to 8 (see \ref{FigaverageN}), the probability $P_{n=8}$ tends to a limit which is less than 0.4 (see \ref{FigprobabilityN} and \ref{FigmaxProb}).

\begin{figure}
  \includegraphics[width=300pt]{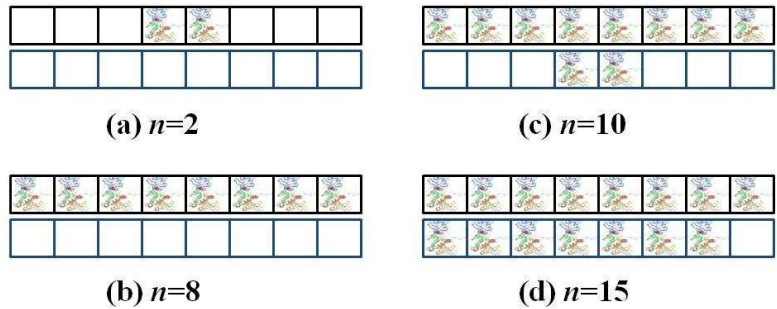}\\
  \caption{Schematic depiction of the most likely configuration of chaperonin TRiC/CCT if it has $n$ ATP bound subunits, for $n=2, 8, 10, 15$. But one should keep in mind that, if the chaperonin is incubated in high [ATP] solutions, it will most likely be bound with $n=8$ ATP molecules [see \ref{FigprobabilityN} and \ref{FigmaxProb}].}\label{FigstateN}
\end{figure}

Since $2\hat\alpha_1<2\hat\beta_1<2\hat\alpha_0$ (see \ref{table1}), if there is only one subunit of the present TRiC/CCT which is ATP bound, the new arrival ATP molecule will most likely attach to one of its adjacent subunits (in the same ring), or for simplicity, denote such process as follows $[010\cdots0][000\cdots0]+1\rightarrow [011\cdots0][000\cdots0]$ or $[110\cdots0][000\cdots0]$. Similarly, one can easily show that, $[01100000][000\cdots0]+1\rightarrow [01110000][000\cdots0]$ or $[11100000][000\cdots0]$ (since $2\hat\alpha_1+\hat\alpha_2<\hat\beta_1+\hat\beta_2+\hat\alpha_1<2\hat\alpha_1+\hat\alpha_0$), and
$[01110000][000\cdots0]+1\rightarrow [01111000][000\cdots0]$ or $[11110000][000\cdots0]$(since $2\hat\alpha_1+2\hat\alpha_2<\hat\alpha_1+\hat\alpha_2+\hat\beta_1+\hat\beta_2
<2\hat\alpha_1+\hat\beta_1+\hat\beta_3$), etc.
One can also verify that for $k\ge 4$, $2\hat\alpha_1+(k-2)\hat\alpha_2
<\hat\alpha_1+(k-3)\hat\alpha_2+\hat\beta_1+\hat\beta_2
<2\hat\alpha_1+(k-4)\hat\alpha_2+\hat\beta_1+\hat\beta_3$. Therefore, if bound ATP number $2\le n\le8$, the most likely state of TRiC/CCT is that, all these $n$ ATP molecules are in the same ring and adjacent to each other. Moreover, one can easily show that for any $2\le n\le 16$, all the bound ATP molecules will most likely gather together (see \ref{FigstateN}). Since $\hat\alpha_2<\hat\alpha_1<\hat\alpha_0$, the ATP binding to the same ring is positively cooperated. However, $\hat\alpha_2<0$, $\gamma_0<0$ but $\hat\beta_i>0$ implies that the ATP binding to different rings is negatively cooperated.

\section{Conclusions}

In conclusion, from the viewpoint of statistical thermodynamics, a simple method to describing the ATP binding process to multisubunit enzymes is present. Each subunit of enzymes might be in empty or ATP bound state, so there are altogether $2^M$ states for one single enzyme with $M$ subunits. The potential of each subunit depends not only on ATP binding to itself but also on ATP binding to its adjacent (and {\it mirror}) subunits. The probability of enzyme in a special state can be obtained by the Boltzmann's distribution. Consequently, the probability of bound ATP number in one single enzyme and the corresponding average value of ATP number can be obtained. This model can fit the recent experimental data for chaperonin TRiC/CCT well \cite{Jiang2011}. We find that,
in saturating ATP solutions, the average number of ATP bound to one single TRiC/CCT tends to 8, and the maximum of probability $P_n$ that there are $n$ ATP bound to one single TRiC/CCT tends to a value less than 0.4. The most likely state of TRiC/CCT is that, all the bound ATP molecules are adjacent (or {\it mirror}) to each other, and for ATP number $2\le n\le 8$, all the bound ATP molecules will most likely be in the same ring. The cooperation of ATP binding to the same ring of TRiC/CCT is positive, but it is negative for ATP binging to the different rings.

%% classes.  Rather than use \section, an appropriate macro is
%% provided that will always work.
%%%%%%%%%%%%%%%%%%%%%%%%%%%%%%%%%%%%%%%%%%%%%%%%%%%%%%%%%%%%%%%%%%%%%
\acknowledgement

This study is funded by the Natural Science Foundation of Shanghai (under Grant No. 11ZR1403700).

%% \suppinfo macro.
%%%%%%%%%%%%%%%%%%%%%%%%%%%%%%%%%%%%%%%%%%%%%%%%%%%%%%%%%%%%%%%%%%%%%

\providecommand*{\mcitethebibliography}{\thebibliography}
\csname @ifundefined\endcsname{endmcitethebibliography}
{\let\endmcitethebibliography\endthebibliography}{}

\newpage

%\onecolumngrid

\centerline{}
\centerline{}
\vskip2cm

%\begin{widetext}

%\end{widetext}

\newpage

\centerline{}
\centerline{}
\vskip5cm

\begin{table}
    \centering
    \caption{Model parameters used in the theoretical plots in  \ref{FigaverageN}, \ref{FigprobabilityN} and \ref{FigmaxProb}. In the calculations, we use $N=N_0=25$ for the chaperonin TRiC/CCT incubated in 5$\mu $M ATP solution, and $N=N_0$[ATP]$/5$ for other ATP solutions. All the following parameter values and $N_0=25$ are obtained by fitting \ref{eq10} \ref{eq12} to the experimental data obtained by Jiang {\it et al} \cite{Jiang2011}. The unit of parameters $\hat \alpha_i, \hat \beta_i$ is  $k_BT$.}
    \begin{tabular}{ccccccc}
      \hline\hline
    $\hat \alpha_0$\ \ & $\hat \alpha_1$ \ \ & $\hat \alpha_2$ \ \ & $\hat \beta_1$ \ \ &  $\hat \beta_2$  \ \ &  $\hat \beta_3$  \ \ & $\gamma_0$ \ \ \\
    \hline
    5.33 & 2.28 & -0.04 & 3.07 & 1.32 & 0.31 & -76.50 \\
    \hline\hline
    \end{tabular}
    \label{table1}
\end{table}

\end{document}